\documentclass[twocolumn,showpacs,prl,10pt,aps,floatfix]{revtex4-1}
\usepackage[english]{babel}
\usepackage{amsmath,amssymb}
\usepackage{times}
\usepackage{epsfig}
\usepackage[colorlinks,linkcolor=blue,citecolor=blue]{hyperref}
\usepackage{color}
\usepackage{mhchem}
\begin{document}
\title{Local spin relaxation within the random Heisenberg chain}
\author{J. Herbrych$^{1}$}
\author{J. Kokalj$^{1}$}
\author{P. Prelov\v{s}ek$^{1,2}$}
\affiliation{$^1$J. Stefan Institute, SI-1000 Ljubljana, Slovenia}
\affiliation{$^2$Faculty of Mathematics and Physics, University of Ljubljana, SI-1000 Ljubljana,
Slovenia}
\date{\today}
\begin{abstract}
Finite--temperature local dynamical spin correlations $S_{nn}(\omega)$ 
are studied numerically within the random
spin--$1/2$ antiferromagnetic Heisenberg chain. The aim is to explain measured NMR spin--lattice
relaxation times in \ce{BaCu_{2}(Si_{0.5}Ge_{0.5})_{2}O_{7}}, which
is the realization of a random spin chain. In agreement with experiments
we find that the distribution of relaxation times within the model shows
a very large span similar to the stretched--exponential
form. The distribution is strongly reduced with increasing $T$, 
but stays finite also in the high--$T$
limit. Anomalous dynamical correlations can be associated to the random 
singlet concept but not directly to static quantities. Our results also 
reveal the crucial role of the spin anisotropy (interaction), since
the behavior is in contrast with the ones for XX model, where we do not find 
any significant $T$ dependence of the distribution. 
\end{abstract}
\maketitle

One--dimensional (1D) quantum spin systems with random exchange
couplings reveal interesting phenomena fundamentally different from
the behavior of ordered chains. Since the seminal studies of
antiferromagnetic (AFM) random Heisenberg chains (RHC) by Dasgupta and
Ma \cite{ma1979,dasgupta1980} using the renormalization--group
approach and further development by Fisher \cite{fisher1994}, it has
been recognized that the quenched disorder of exchange couplings $J$
leads at lowest energies to the formation of random singlets with
vanishing effective $\tilde J$ at large distances. The consequence
for the uniform static susceptibility $\chi^0$ is the singular Curie--type
temperature ($T$) dependence, dominated by nearly uncoupled spins at
low--$T$ and confirmed by numerical studies of model systems
\cite{hirsch1980}, as well by measurements of $\chi^0(T)$ on the class
of materials being the realizations of RHC physics, in particular the
mixed system \ce{BaCu_{2}(Si_{1-x}Ge_{x})_{2}O_{7}}
\cite{zheludev2007,shiroka2011,shiroka2012}.

Recent measurements of NMR spin--lattice relaxation times $T_1$ in
\ce{BaCu_{2}(Si_{0.5}Ge_{0.5})_{2}O_{7}} \cite{shiroka2011}
reveal a broad distribution of different $T_1$ resulting in a
nonexponential magnetization decay being rather of a
stretched--exponential form. In connection to this
the most remarkable is the strong
$T$ dependence of the $T_1$ span becoming progressively large and the
corresponding distribution non--Gaussian at low--$T$. It is evident that in a
random system $T_1$, which is predominantly testing the local spin correlation
function $S_{nn}(\omega \to 0)$, becomes site $n$ dependent and we are therefore
dealing with the distribution of $T_{1n}$ leading to a nonexponential magnetization
decay.

Theoretically the behavior of dynamical spin correlations in RHC 
has not been adequately addressed so far. 
There is (to our knowledge) no established model result 
and moreover no clear prediction for the behavior of dynamical ($\omega \ne 0$) 
spin correlations at $T>0$ in RHC.
It seems plausible that the low--$T$
behavior should follow from the random--singlet concept and its
scaling properties, discussed within the framework of the
renormalization--group approaches
\cite{dasgupta1980,fisher1994,westerberg1997,motrunich2001}. 
Still, the relation to singular static correlations as evidenced, e.g., by 
$\chi^0(T )$ diverging at $T\to 0$, and low-
$\omega$ dynamical correlations is far from clear.

One open question is also the qualitative similarity to the
behavior of the random anisotropic XX chain invoked in several studies
\cite{bulaevskii1972,hirsch1980,westerberg1997,motrunich2001}. The
latter system is equivalent to more elaborated problem of
noninteracting (NI) spinless fermions with the off--diagonal (hopping)
disorder \cite{theodorou1976,eggarter1978}.

In the following we present results for the dynamical local spin
correlation function $S_{nn}(\omega)$, in particular for its limit 
$s=S_{nn}(\omega \to 0)$ relevant for the NMR $T_1$, 
within the AFM RHC
model for $T>0$, obtained using the numerical method based on the
density--matrix renormalization group (DMRG) approach
\cite{kokalj2009}. At high $T \geq J$, distribution of $s$
reveals a modest but finite width
qualitatively similar both for the isotropic and the XX chain. On the
other hand, the low--$T$ variation established numerically is essentially
different. While for the XX chain there is no significant $T$
dependence, results for the isotropic case reveal at low $T \ll J$
a very large span of $s$ values and corresponding $T_{1n}$,
qualitatively and even quantitatively
consistent with NMR experiments \cite{shiroka2011}.

We study in the following the 1D spin--$1/2$ model
representing the AFM RHC,
\begin{equation}
H=\sum_{i} J_i\left(S^x_i S^x_{i+1}+S^y_i S^y_{i+1}+\Delta S^z_i S^z_{i+1} \right)\,,
\label{ham}
\end{equation}
where $J_i$ are random and we will assume their distribution as
uncorrelated and uniform in the interval $J-\delta J\le J_i\le
J+\delta J$, with the width $\delta J< J$ as the parameter. In the
following we will consider predominantly the isotropic case $\Delta=1$, but as
well the anisotropic XX case with $\Delta=0$. The chain is of the length $L$ with
open boundary conditions (o.b.c.) as useful for the DMRG method. We
further on use $J=1$ as the unit of energy as well as $\hbar=k_B=1$.

Our aim is to analyse the local spin dynamics in connection with the
NMR spin--lattice relaxation \cite{shiroka2011}. In a homogeneous
system the corresponding relaxation rate $1/T_1$ is expressed in terms of the
$q$--dependent spin correlation function,
\begin{equation}
\frac{1}{T_{1}}=\sum_{q \alpha } A^2_\alpha(q) S^{\alpha \alpha}(q,\omega\to 0)\,,
\label{st1}
\end{equation}
where $A^2_\alpha(q)$ involve hyperfine interactions and NMR form
factors \cite{shiroka2011}. 
In the Supplement \cite{supp} we show that the dominant dynamical 
$\omega \to 0$ contribution at
low--$T$ is coming from the regime $q \sim \pi$. Therefore
the variation $A^2_\alpha(q)$ is not essential and the rate
depends only on the local spin correlation function $1/T_{1} \propto
S^{zz}_{\mathrm{loc}}(\omega \to 0)$. In a system with quenched
disorder the relaxation time becomes site dependent, i.e. $T_{1n}$, hence
we study in the following the local correlations $S_{nn}(\omega)$ and the 
distribution of local limits $s=S_{nn}(\omega\to 0)$ and related relaxation
times $\tau=1/s$ where
\begin{equation}
S_{nn}(\omega)=\frac{1}{\pi}\mathrm{Re}\int\limits_{0}^{\infty}\mathrm{d}t\,
{\rm e}^{\imath \omega t}\langle S^z_n(t)S^z_n(0)\rangle\,.
\label{auco}
\end{equation}

In order to reduce finite--size effects we study large systems employing
the finite--temperature dynamical DMRG (FTD--DMRG)
\cite{kokalj2009,schollwock2005,prelovsek2013} method to evaluate the dynamical
$S_{nn}(\omega)$, Eq.~\eqref{auco}.
To reduce edge effects we choose the local site $n$ to be in the middle of the chain,
$n=L/2$. The distribution of $s$ is then calculated with $N_r \sim 10^3$ different
realizations of the system with random $J_i$.
More technical detail on the
calculation can be found in the Supplement \cite{supp}.

\begin{figure}[!ht]
\includegraphics[width=1.0\columnwidth]{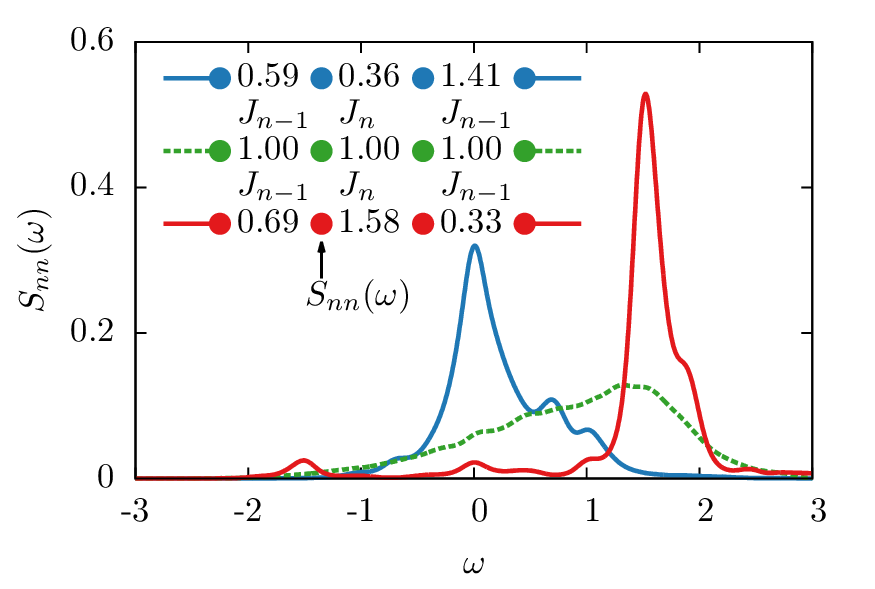}
\caption{(Color online) Dynamical local spin correlations
 $S_{nn}(\omega)$ for different configurations of $J_i$. 
 Shown are spectra for the homogeneous case
 $\delta J=0$ and two configurations with $\delta J=0.7$, calculated
 for $T=0.5$ and $L=80$ sites.}
\label{spect}
\end{figure}

We start the presentation of results with typical examples of
$S_{nn}(\omega)$. In Fig.~\ref{spect} we show calculated spectra for
system with $L=80$ sites, $T=0.5$, $\Delta=1$, and three different
realizations of $J_i$, i.e. the homogeneous system with $J_i=1$ and two
configurations with $\delta J= 0.7$.
Spectra for the uniform
system are broad and regular at $\omega\sim 0$ agreeing with those
obtained with other methods \cite{naef1999}, while
$S_{nn}(\omega)$ for random case strongly depend apart from $\delta J$
also on the local $J_i, i \sim n$.
In particular, spectra
with both $J_{n-1}$ and $J_{n}$ small have large amplitude at the
relevant $\omega \sim 0$, while spectra with one large $J_{n-1}$ or
$J_{n}$ have most of the weight at high--$\omega$ and small
amplitude at $\omega \sim 0$ (elaborated further in the
conclusions).
For the following analysis it is important that
$s=S_{nn}(\omega \to 0)$ can be extracted reliably.

\begin{figure}[!ht]
\includegraphics[width=1.0\columnwidth]{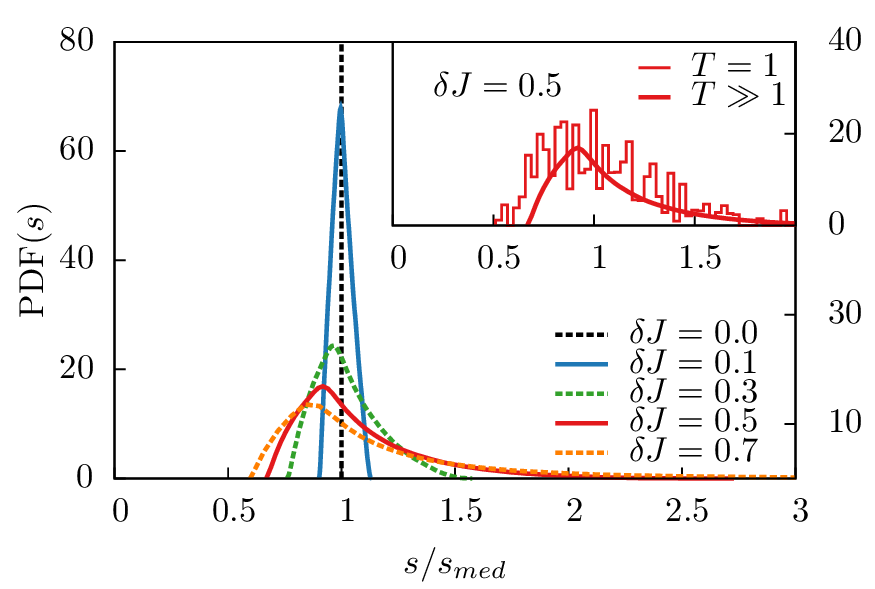}
\caption{(Color online) Probability distribution function of local relaxation rates PDF$(s)$ at
$T\gg 1$ evaluated using the moment expansion for different $\delta J$. Inset: Comparison of
analytical and FTD--DMRG result for $\delta J=0.5$, $L=20$ with
full basis and averaged over $N_{r}=10^3$ realizations.}
\label{momex}
\end{figure}

{\it Results for $T \geq J$:} Before displaying results for most
interesting $T<J$ regime, we note that even at $T \gg J$ one cannot
expect a well defined $\tau=\tau_0$ but rather a distribution
of values. One can understand this by studying analytically local frequency moments 
within the high--$T$ expansion and using the Mori's continued fraction
representation \cite{mori1965} with the Gaussian--type truncation at
the level of $l>3$ \cite{tommet1975,oitmaa1984} (see \cite{supp}
for more details).
In the inset of Fig.~\ref{momex} we present the high--$T$
result for $\mathrm{PDF}(s)$ and
compare it with the numerical results evaluated for $T=1$.
Several conclusions can be drawn from results presented on Fig.~\ref{momex}: (a) The
agreement of PDF$(s)$ obtained via the analytical approach and
numerical FTD--DMRG method is satisfactory having the origin in quite
broad and featureless spectra $S_{nn}(\omega)$ at $T \geq J$. Still we note
that median value of $s$ ($s_{\mathrm{med}}$) differ between both approaches
and that for $T\gg J$ (unlike $T \leq J$)
contribution of $q\to 0$ can become essential \cite{supp,sandvik1997}. (b)
PDF$(s)$ becomes quite asymmetric and broad for $\delta J\ge 0.5$. (c)
Consequently, also the distribution of local relaxation times
PDF$(\tau)$ has finite but modest width for $T \to \infty$. This seems in
a qualitative agreement with NMR data for
\ce{BaCu_{2}(Si_{0.5}Ge_{0.5})_{2}O_{7}},
where the width was hardly detected at
high--$T$ \cite{shiroka2011}.

\begin{figure}[!ht]
\includegraphics[width=1.0\columnwidth]{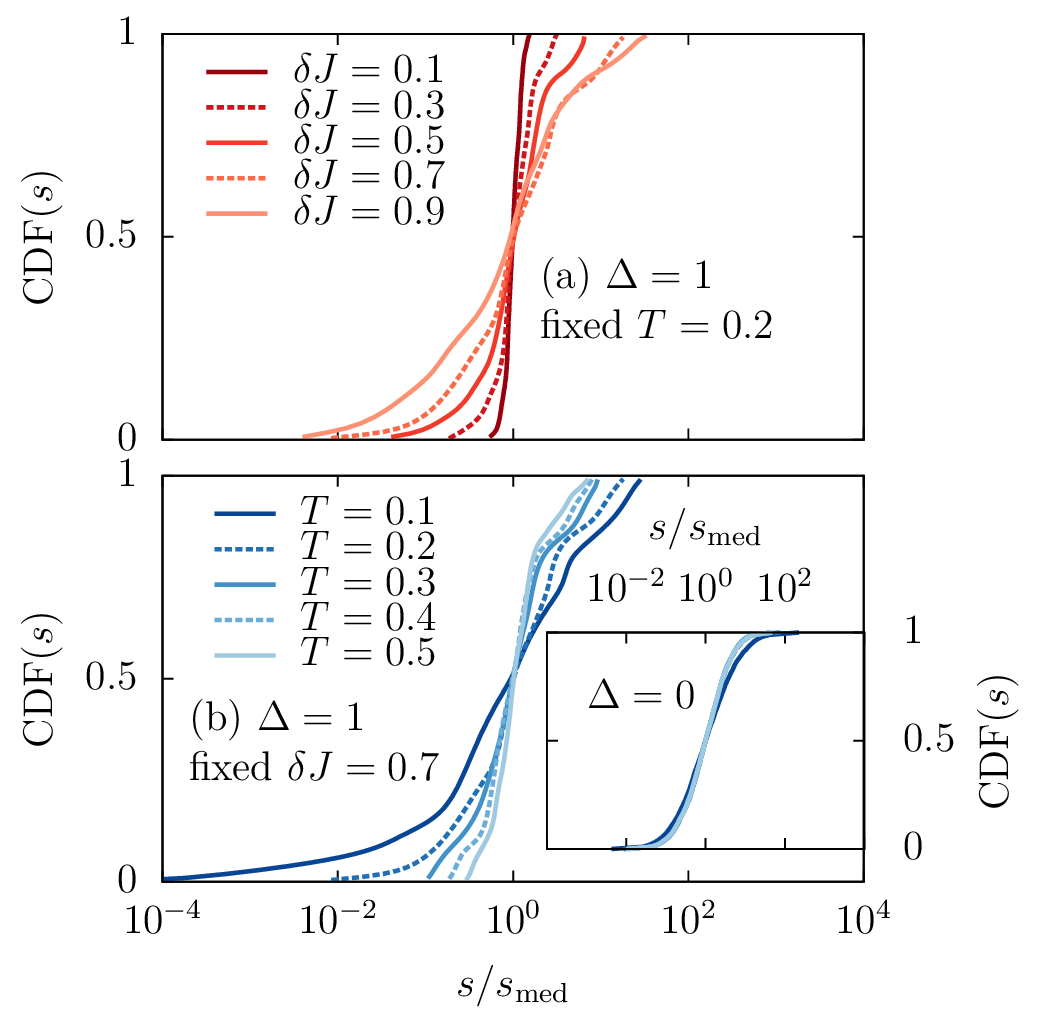}
\caption{(Color online) Cumulative distribution function of $s$. Shown are FTD--DMRG results
 for $\Delta=1$: (a) for fixed $T=0.2$ and various $\delta J$, (b)
 for fixed $\delta J=0.7$ and various $T \leq 0.5 $.
 Inset of (b):full diagonalization results for $\Delta=0$, $\delta J=0.7$ and various $T$.}
\label{cdf}
\end{figure}

{\it Results for $T < J$:} More challenging is the
low--$T$ regime which we study using the FTD--DMRG method for
typically $L=80$ and $N_r\sim 10^3$. Besides the
isotropic case $(\Delta =1 )$, we investigate for comparison also the
XX model ($\Delta=0$). As the model of NI fermions with the
off--diagonal disorder \cite{bulaevskii1972,theodorou1976} it can be
easily studied via full diagonalization on much longer chains with
$L \sim 16000$. PDF for $T<J$ can become very broad and
asymmetric. Hence, we rather present results as the cumulative distribution
$\mathrm{CDF}(x)=\int_{0}^{x}\mathrm{d}y\,\mathrm{PDF}(y)$. Further we
rescale $x$ values to the median defined as
$\mathrm{CDF}(x_{\mathrm{med}})=0.5$. Results for
CDF$(s)$ are presented in Fig.~\ref{cdf}. Note that
$\mathrm{PDF}(\tau)=\mathrm{PDF}(s)/\tau^{2}$.
Panels in Fig.~\ref{cdf} represent results for the isotropic case $\Delta=1$ with
(a) fixed $T=0.2$ and varying $\delta J=0.1-0.9$, while in (b) $\delta
J=0.7$ is fixed and $T=0.1 - 0.5$.
Inset of Fig.~\ref{cdf}b displays
the $T$ dependence (for fixed $\delta J=0.7$) of CDF for the XX chain.

We first note that within the XX chain CDF$(s)$ are
essentially $T$ independent. This appears as quite a contrast to, e.g.,
static $\chi^0(T)$ which exhibits a divergence at $T \to 0$
\cite{hirsch1980,supp}. Results for the isotropic case $\Delta=1$ in
Figs.~\ref{cdf}a,b are evidently different. The span in CDF becomes
very large (note the logarithmic scale) either by increasing $\delta
J$ at fixed $T$ or even more by decreasing $T$ at fixed $\delta
J$. From the corresponding PDF one can calculate the relaxation
function $R(t)=\int \mathrm{d}s\,\mathrm{PDF}(s){\rm e}^{-ts}$, which
is in fact the quantity measured in the NMR as a 
time--dependent magnetization recovery \cite{shiroka2011}. As in
experiment the large span in our results for low--$T$ can be
captured by a stretched exponential form,
$R(t)\approx\exp[-\left(t/\tau_0 \right)^{\Gamma}]$,
where $\Gamma$ and $\tau_0$ are parameters to be fitted for particular
PDF$(s)$ and corresponding $R(t)$. It is evident that $\Gamma \ll 1$ means
large deviations from
the Gaussian--like form, and in particular very pronounced tails in
PDF$(s)$, both for $s \gg s_{\mathrm{med}}$ as well as a singular
variation for $s \to 0$. In the latter
regime $1/\tau_0$ can deviate substantially from average of local
$1/\tau$. It should be also noted that stretched exponential form,
is the simplest one capturing the large span of $s$ values. It is also used 
in the experimental analysis \cite{shiroka2011}, but the corresponding 
PDF$(s)$ reveal somewhat enhanced tails for $s > s_{\mathrm{med}}$
relative to calculated ones in Fig.~\ref{cdf}a,b, and the opposite trend for 
$s < s_{\mathrm{med}}$. This suggests possible improvements and 
description beyond stretched exponential form, which we leave as a
future challenge. More details can be found in the Supplement \cite{supp}.

\begin{figure}[!ht]
\includegraphics[width=1.0\columnwidth]{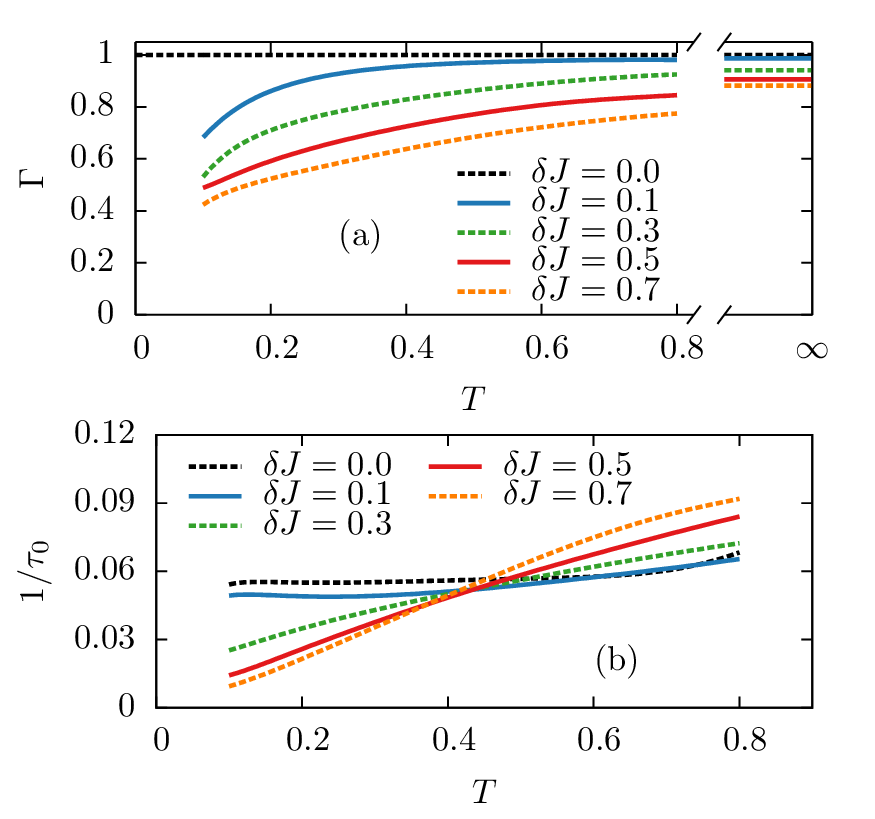}
\caption{(Color online) (a) Exponent $\Gamma$ vs. $T$ obtained from PDF$(s)$ data for different
$\delta J$ and isotropic case $\Delta=1$. (b) $T$ dependence of fitted $1/\tau_0$ for $\Delta=1$
and different $\delta J$.}
\label{fit}
\end{figure}

Results for the fitted exponent $\Gamma(T)$ for $\Delta=1$ as
extracted from numerical PDF$(s)$ for various $\delta J$ are shown in
Fig.~\ref{fit}a. They confirm experimental observation
\cite{shiroka2011} of increasing deviations from simple exponential
variation ($\Gamma=1$) for $T \ll J$. While for $T > J$, $\Gamma
\lesssim 1$ for modest $\delta J < 0.7$, low--$T$ values can reach
even $\Gamma < 0.5$ at lowest reachable $T < 0.1$. Note that in such a
case values of $s$ are distributed over several orders of magnitude.

Of interest for the comparison with experiment is also the
$T$ variation of fitted $1/\tau_0$. Results are again essentially
different for $\Delta=0$ and $\Delta=1$. $\tau_0$ (as well
$s_{\mathrm{med}}$) for $\Delta=0$ follows well the Korringa law
$1/\tau_0 \propto T$ for
$T<0.5$, as usual for the system of NI fermions with a constant density of states (DOS)
(divergent DOS at $E \to 0$ could induce a logarithmic correction). 
On the other hand, for the isotropic ($\Delta=1$)
chain with no randomness $\tau_0=\tau$ it should follow $1/\tau
\sim \mathrm{const.}$ for $T <J$ \cite{sandvik1995,naef1999}. Similar
behavior is observed for weak disorder $\delta J=0.1$ as shown in
Fig.~\ref{fit}b. However, with increasing randomness $\delta J$,
$1/\tau_0$ becomes more $T$ dependent and increases with $T$.
Such $T$ dependence in the RHC of $1/\tau_0$ is, although in
agreement with experiment, in apparent contrast with diverging
$\chi^0(T\to 0)$. This remarkable dichotomy between static
and dynamical $\omega \to 0$ behaviour can be reconciled 
by the observation that in a random system $S(q,\omega\sim 0)$ reveals 
besides the regular part also a delta peak at $\omega=0$ (not entering $1/T_1$), 
which can be traced back to diagonal matrix elements \cite{supp} being 
an indication of a nonergodic behaviour (at least at low--$T$). 
Note that more frequently studied static $S(q)$ (equal-time correlation) 
\cite{hoyos2007} represents a sum rule containing both parts. Also, the relation
$\chi^0(T)=S(q=0)/T$ in spite of divergent $\chi(T \to 0)$ leads to 
vanishing $S(q=0)$  at $T\to 0$ only slower than linearly \cite{supp,hoyos2007}.

As a partial summary of our results, we comment on the relation to the experiment
on \ce{BaCu_{2}(Si_{0.5}Ge_{0.5})_{2}O_{7}} \cite{zheludev2007,shiroka2011}.
The spin chain is in this
case assumed to be random mixture of two different values $J_i=280$~K,
$580$~K, which correspond roughly to our $\delta J\simeq0.6$ (fixing the
same effective width) and $J=430$~K. Taking these values, our results
for $\Gamma(T)$ 
as well as $1/\tau_0(T)$ agree well with experiment. In particular we note that at lowest $T \ll
J$ our calculated $\Gamma \sim 0.5$ for $\delta J=0.6$ matches the
measured one. Some discrepancy appears to be a steeper increase of
measured $\Gamma(T)$ towards the limiting $\Gamma=1$ coinciding with
observed very narrow PDF$(\tau)$ which remains of finite width in our
results even for $T \to \infty$ as seen in Fig.~\ref{momex}. 
As far as calculated $1/\tau_0(T)$ vs. NMR experiment is concerned we note 
that taken into account the normalization of average $J$
disordered system reveals at $T \to 0$ smaller $1/\tau_0$ than a pure one consistent with 
the experiment \cite{shiroka2011}. In agreement with the experimental analysis is
also strong $T$ variation of $1/\tau_0$ at low--$T$ in disordered system in
contrast to a pure one. 

Our results on the local spin relaxation $S_{nn}(\omega)$ and in
particular its $T$ dependence cannot be directly explained within the
framework of existing theoretical studies and scaling approaches to
RHC \cite{dasgupta1980,fisher1994,motrunich2001}.
Our study clearly
shows the qualitative difference in the behavior of the XX chain and
the isotropic RHC. While in the former model mapped on NI electrons,
$T$ does not play any significant role on PDF$(s)$ as seen in inset of
Fig.~\ref{cdf}b, $\Delta=1$ case shows strong variation with $T \ll
J$. It is plausible that the difference comes from the interaction and
many--body character involved in the isotropic RHC. To account for
that we design in the following a simple qualitative argument.

The behavior of $S_{nn}(\omega\sim 0)$ at low--$T$ is dominated by
transitions between low--lying singlet and triplet states which become
in a RHC nearly degenerate following the scaling arguments with
effective coupling $\tilde J \to 0$ for more distant spins and
reflected in diverging $\chi^0(T \to 0)$
\cite{supp,dasgupta1980,hirsch1980,fisher1994}. Such transitions are
relevant at $\omega \to 0$ behavior as presented in
Fig.~\ref{spect}. Moreover, local $S_{nn}(\omega \sim 0)$ exhibit 
large spread due to the variations in the local environment. Let us for simplicity
consider the symmetric Heisenberg model on four sites (with o.b.c.)
with a stronger central bond $J_2 \gg J_1=J_3$ and
$J=(J_1+J_2+J_3)/3$. It is then straightforward to show that the
lowest singlet--triplet splitting is strongly reduced, i.e. $\Delta E
\propto \eta^2 J$ where $\eta=J_1/J_2 $. Within the same model one can
evaluate also the ratio between two different amplitudes of
$S_{nn}(\omega \sim \Delta E)=A_{nn}\delta(\omega-\Delta E)$, on sites
$n=1,2$ neighboring the weak and strong bond,
\begin{equation}
\frac{1}{W}= \frac{A_{22}}{A_{11}}= \frac{ |\langle \Psi_t |S^z_2|\Psi_s\rangle |^2}
{|\langle \Psi_t |S^z_1|\Psi_s\rangle |^2}\sim \eta^2\,.
\label{3s}
\end{equation}
The relation shows that the span between largest and smallest
amplitudes increases as $W\propto 1/\eta^2 \propto 1/\Delta
E$. Continuing in the same manner the scaling procedure for AFM RHC
\cite{dasgupta1980,fisher1994} for a long chain the smallest effective
coupling between further spins $\tilde J$ vanishes at $T=0$ and $\Delta
E \propto \tilde J \to 0$, so that one expects $W \to \infty$ for $T \to
0$. On the other hand, for $T>0$ the scaling should be cut off at
$\tilde J \sim T$ at least for $\Delta=1$, finally leading to the
strong $W(T)$ dependence ($W \propto 1/T$). 

In the summary, we have
reproduced qualitatively main experimental NMR 
results on mixed system \ce{BaCu_{2}(Si_{0.5}Ge_{0.5})_{2}O_{7}} including
anomalously wide distribution of relaxation rates, together with $T$
dependencies of experimental parameters ($1/\tau_0$, $\Gamma$) and
provide microscopic explanation with the help of the random-singlet
framework. Our qualitative conclusions on the RHC do not change by
changing $S^z_{\mathrm{tot}}$ (adding finite field in the fermionic language)
or even reducing $\Delta <1$ provided that $\Delta > 0$ (see Supplement \cite{supp}).
We also comment on striking difference between static and dynamic quantities
and observed deviations from stretched exponential phenomenology.

\acknowledgements{We would like to thank M. Klanj\v{s}ek for useful discussions. This research
was supported by the RTN-LOTHERM project and the Slovenian Agency grant No. P1-0044.}


\clearpage
\appendix
\setcounter{figure}{0}
\setcounter{equation}{0}
\newcommand{\rom}[1]{\uppercase\expandafter{\romannumeral #1\relax}}
\renewcommand{\citenumfont}[1]{S#1}
\renewcommand{\bibnumfmt}[1]{[S#1]}
\renewcommand{\thefigure}{S\arabic{figure}}
\renewcommand{\theequation}{S\arabic{equation}}

\begin{center}
\bf{\uppercase{Supplementary Material} for ``Local spin relaxation within random Heisenberg
chain''}
\end{center}

\section{\rom{1.} \uppercase{Numerical method}}
In this section we present in more detail the numerical method,
finite--temperature dynamical DMRG (FTD--DMRG). The method is a
variation of a zero temperature ($T=0$) DMRG
\cite{s_white1992, s_schollwock2005}, with targeting
of the ground state or ground state density matrix
$\rho_0=|0\rangle\langle 0|$ generalized to targeting of the
finite--$T$ density matrix $\rho^\beta= \frac{1}{Z}\sum_n |n\rangle
\textrm{e}^{-\beta H}\langle n|$,
\cite{s_kokalj2009,s_kokalj2010,s_prelovsek2011}. Similar
generalization is applied to targeting of the operator on the ground
state. From such targets, the reduced
density matrix is calculated and then truncated in the standard DMRG
like manner for basis optimization. All quantities, that need to be
evaluated at finite--$T$, are calculated with the use of
finite--temperature Lanczos method (FTLM)
\cite{s_jaklic2000,s_prelovsek2011}, which in FTD--DMRG replaces $T=0$
Lanczos method used in the standard DMRG algorithm.

The method is most efficient at low--$T$ and for low frequencies,
where basis can be efficiently truncated and only small portion ($M$
basis states) of the whole basis for block can be kept. In this regime large
system sizes can be reached. The truncation error becomes larger at
higher--$T$, and one needs to either use larger $M$, or reduce system
size, which is legitimate approach, since finite size effects are
smaller at higher--$T$ due to reduced correlation lengths.

We typically keep $M \sim 200$ basis states in the DMRG block and use
systems with length $L \sim 80$ at low $T < J/2$, while for $T > J/2$
smaller systems are employed down to $L \sim 20$, for which full basis
can be used. We stress that randomness of $J_i$ reduces the truncation
error since some larger values of $J_i$ induce strong tension for
formation of a local singlet and therefore in turn reduces the
entanglement on larger distances. Also the local operator, acting on
the middle of the chain, where the local one site basis is not truncated, helps
in this respect.

\begin{figure}[!ht]
\includegraphics[width=0.75\columnwidth]{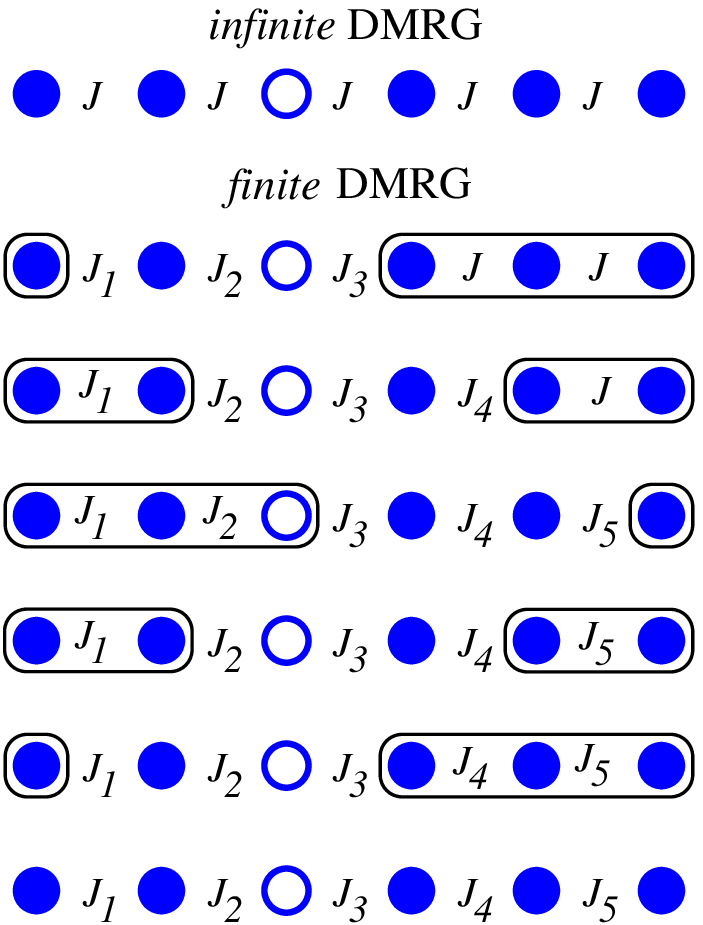}
\caption{(Color online) Schematic ($L=6$) representation of the
beginning of the sweeping method in {\it finite}--DMRG
algorithm, in which randomness is introduced. Open circle represents
site of a local operator used to calculated local spin correlation function
in the {\it measurement} part of the DMRG method.}
\label{sche}
\end{figure}

The quenched random $J_i$ are introduced into the DMRG procedure at
the beginning of {\it finite} algorithm. {\it Infinite} algorithm is
preformed for homogeneous system $J_i=J$ and the randomness of $J_i$
is introduced in the first sweep (see Fig.~\ref{sche} for schematic
presentation). In this way the preparation of the basis in the {\it
 infinite} algorithm is performed just once and for all realizations
of $J_i$--s, while larger number of sweeps (usually $\sim 5$) is
needed to converge the basis within the {\it finite} algorithm for
random $J_i$. After {\it finite} algorithm local dynamical spin
structure factor $S_{nn}(\omega)$ at desired $T$ is calculated for the
site in the middle of the chain within {\it measurements} part of DMRG
procedure.

Any spectra on finite system consists of separate $\delta$--peaks,
which we broaden by changing them into Gaussian with
small broadening $\delta=0.05$ and in this way obtain a smooth
spectra.

Since NMR relaxation rate is related to $S_{nn}(\omega \to 0)$, we are
interested in the limit $\omega \to 0$, which should be contrasted
with the singular $S_{nn}(\omega=0)$.  In order to avoid the problem of diagonal elements and
keeping $\omega \neq 0$ we perform the evaluation of $S_{nn}(\omega
\to 0)$ in the magnetization sector $S^z_{\mathrm{tot}}=0$, which in
terms of spinless fermions corresponds to the canonical ensemble (in
the thermodynamic limit canonical and grand canonical give the same result)
and we remove $\delta(\omega=0)$ peak (see also Section~\rom{4}).

\begin{figure}[!ht]
\includegraphics[width=1.0\columnwidth]{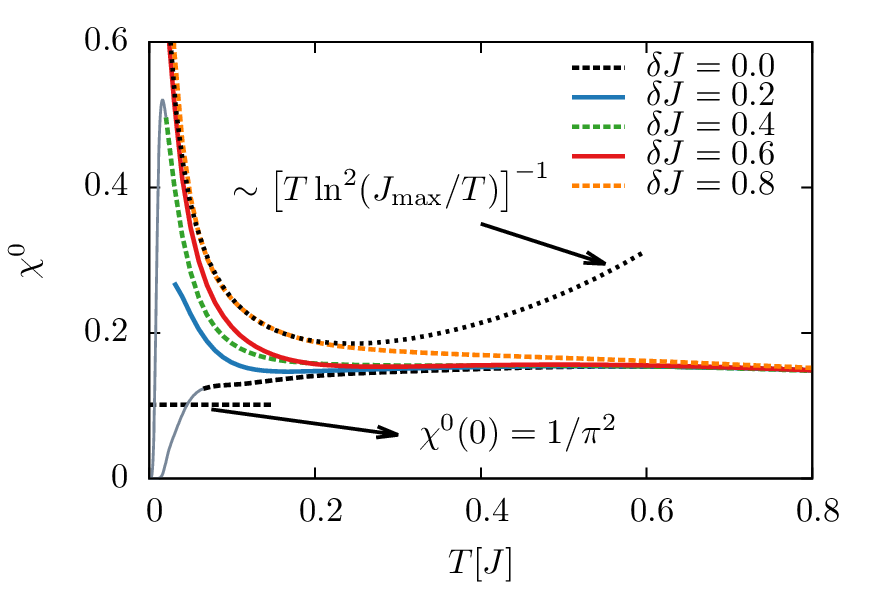}
\caption{(Color online) Static spin susceptibility $\chi^0(T)$
vs. temperature $T$ for various randomness $\delta J$. For random
system ($\delta J \neq 0$) $\chi^0(T)$ is strongly increased at
low--$T$ and agrees with the random singlet \cite{s_fisher1994}
prediction (black dotted line). Sudden drop of $\chi^0(T)$ at
low--$T$ shown for $\delta J=0$ and $\delta J=0.4$ represents opening of
finite--size gap, which is strongly reduced in the random (shown for $\delta
J=0.4$) system. Results are obtained with the finite--temperature
Lanczos method \cite{s_jaklic2000,s_prelovsek2011} on $L=24$ sites.}
\label{chi}
\end{figure}

Diagonal elements are however essential when evaluating static uniform
susceptibility $\chi^0(T)=\langle(S^z_{\mathrm{tot}})^2 \rangle/(LT)$,
which we show in Fig.~\ref{chi}. It has been argued
\cite{s_hirsch1980,s_fisher1994,s_zheludev2007}, that in 1D random
Heisenberg chain the density of low lying excitation is strongly
increased, which is observed in diverging $\chi^0(T)$ for $T\to
0$. Our numerical results show similar behavior (see Fig.~\ref{chi}),
which agrees also with experiment \cite{s_shiroka2011,s_masuda2004}.

Increased number of low--lying excitations (see Fig.~\ref{chi})
also reduces the finite size effect, since, e.g., finite size gap is
reduced, and in this way also the temperature $T_\textrm{fs}$, below
which the finite size effects become important. Therefore, smaller $T$
can be numerically reached in a random system.

\section{\rom{2.} \uppercase{High--$T$ expansion}}
The local spin correlation function can be related to the (local) dynamical
spin susceptibility by relation
\begin{equation}
S_{nn}(\omega)\left[1-\exp(-\beta\omega)\right]=\chi_{nn}''(\omega)\,,
\label{fdt}
\end{equation}
with
\begin{equation}
\chi_{nn}(\omega)=\imath\int\limits_{0}^{\infty}\mathrm{d}t\,
\mathrm{e}^{\imath\omega t}\langle[S^z_n(t),S^z_n(0)]\rangle\,.
\end{equation}
Taking the high--$T$ limit ($\beta\to0$) of Eq.~\eqref{fdt} one gets
$\beta S_{nn}(\omega)=\chi_{nn}''(\omega)/\omega$,
which is so--called relaxation function - symmetric with
respect to $\omega=0$, non--negative function. Note that due to symmetric form of
relaxation function all odd frequency moments, $m_{ln}$, are equal to zero.

The local spin correlation function can by expressed by the Mori's continued fraction
representation \cite{s_mori1965}:
\begin{equation}
\hat{S}_{nn}(z=\imath\omega)=
\cfrac{\delta_{0n}}{z+
\cfrac{\delta_{1n}}{z+
\cfrac{\delta_{2n}}{z+
\cdots}}}\,,
\label{moricf}
\end{equation}
where coefficient $\delta_{ln}$ are cumulants of $S_{nn}(\omega)$,
i.e. $\delta_{0n}=m_{0n}$, $\delta_{1n}=m_{2n}/m_{0n}$,
$\delta_{2n}=m_{4n}/m_{2n}-m_{2n}/m_{0n}$. 
$m_{ln}$ are frequency moments of the local spectra, $m_{ln}=\int\mathrm{d}\omega\,\omega^{l}S_{nn}(\omega)$. 

For $l>3$ we chose a truncation $\zeta_n=\delta_{3n}/(z+\dots)$,
which assumes \cite{s_tommet1975,s_oitmaa1984} a Gaussian--like decay
of correlation function,
i.e. $\zeta_n=\sqrt{2/\pi}(\delta_{1n}+\delta_{2n})/\delta_{2n}^{3/2}$.
The $S_{nn}(\omega)$ can be recovered from Eq.~\eqref{moricf} by the relation
$S_{nn}(\omega)=\mathrm{Re}[\hat{S}_{nn}(z=\imath\omega)]/\pi$, leading to
\begin{equation}
S_{nn}(\omega)=\frac{1}{\pi}\frac{\zeta_{n}\delta_{0n}\delta_{1n}\delta_{2n}}
{\left[\omega\zeta_{n}\left(\omega^{2}-\delta_{1n}-\delta_{2n}\right)\right]^2
+\left(\omega^{2}-\delta_{1n}\right)^2}\,.
\label{hislo}
\end{equation}
Note that Eq.~\eqref{hislo} gives the first three nonzero ($l=0,2,4$)
frequency moments
$m_{ln}$ correctly,
independent of a choice of $\zeta_n$.

Frequency moments $m_{ln}$ of $S_{nn}(\omega)$ can be evaluated
analytically for $T=\infty$, e.g., $m_{0n}=\langle
S^{z}_{n}S^{z}_{n}\rangle$, $m_{2n}=\langle
[H,S^{z}_{n}][H,S^{z}_{n}]\rangle$, etc. For zero magnetization,
$S^{z}_{\mathrm{tot}}=(L_{\uparrow}-L_{\downarrow})/2L=0$, where
$L_{\uparrow}$ ($L_{\downarrow}$) is number of up (down) spins, the
first three nonzero moments of the order of ${\cal O}(\beta)$ are:
\begin{eqnarray}
m_{0n}=\frac{1}{4}\,,\qquad
m_{2n}=\frac{J^2_{n-1}+J^2_{n}}{8}\,,
\nonumber\\
m_{4n}=\frac{1+\Delta^2}{32}\left(J^{2}_{n-2}J^{2}_{n-1}+J^{2}_{n}J^{2}_{n+1}\right)
\nonumber\\
+\frac{3+2\Delta^2}{32}\left(J^{4}_{n-1}+J^{4}_{n}\right)
+\frac{7+2\Delta^2}{32}J^{2}_{n-1}J^{2}_{n}\,.
\end{eqnarray}

\begin{figure}[!ht]
\includegraphics[width=1.0\columnwidth]{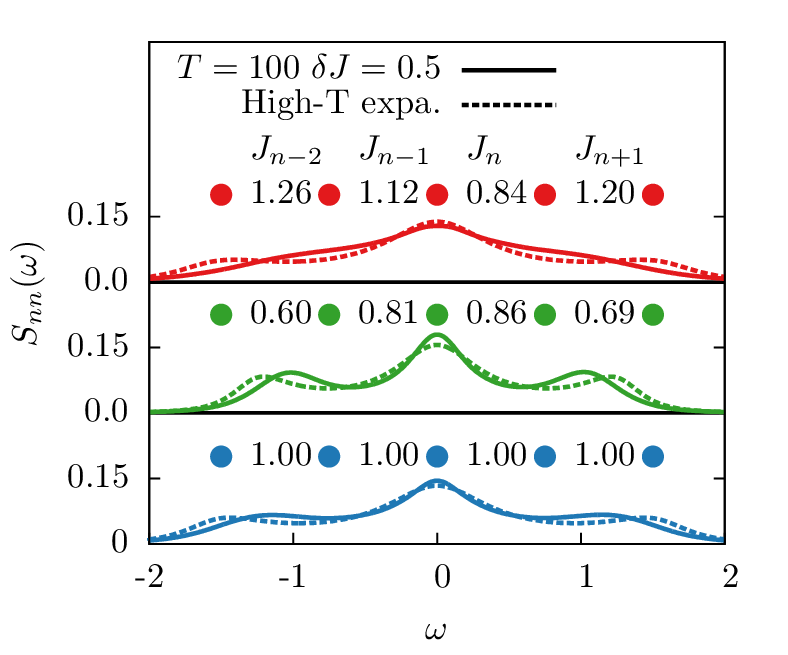}
\caption{(Color online) Comparison of $S_{nn}(\omega)$ between analytical high--$T$ expansion result
and numerical FTD--DMRG result ($L=20$, $T=100$, full basis) for three realization of $J_i$.}
\label{spechT}
\end{figure}

In Fig.~\ref{spechT} we present comparison of high--$T$ expansion
result and FTD--DMRG result ($L=20$, $T=100$, full basis) for $S_{nn}(\omega)$ and
three realizations of $J_i$. One can see that the agreement is
good for actual finite size system. It should be, however, noted that $q \to 0$
contribution (leading to finite size corrections) can become essential for
$T\gg J$ \cite{s_sandvik1997}.

As a final remark of this section we comment on the probability
distribution function (PDF) of $s=S_{nn}(\omega\to 0)$ presented in
Fig.~2 in the main text. Assuming the uniform distribution of $J_i$,
$i=n-2,\cdots,n+1$ the PDF$(s)$ can be can be generated from
expression
\begin{equation}
s=S_{nn}(\omega\to0)=
\frac{1}{\sqrt{8\pi^3}}\frac{\delta_{1n}+\delta_{2n}}{\delta_{1n}\delta_{2n}^{1/2}}\,.
\end{equation}
The PDF-s presented in Fig.~2 (main text) where obtained from
$N_r=10^6$ realizations of $J_i$.

\section{\rom{3.} \uppercase{Finite magnetic field and $\Delta=0.5$}}

\begin{figure}[!ht]
\includegraphics[width=1.0\columnwidth]{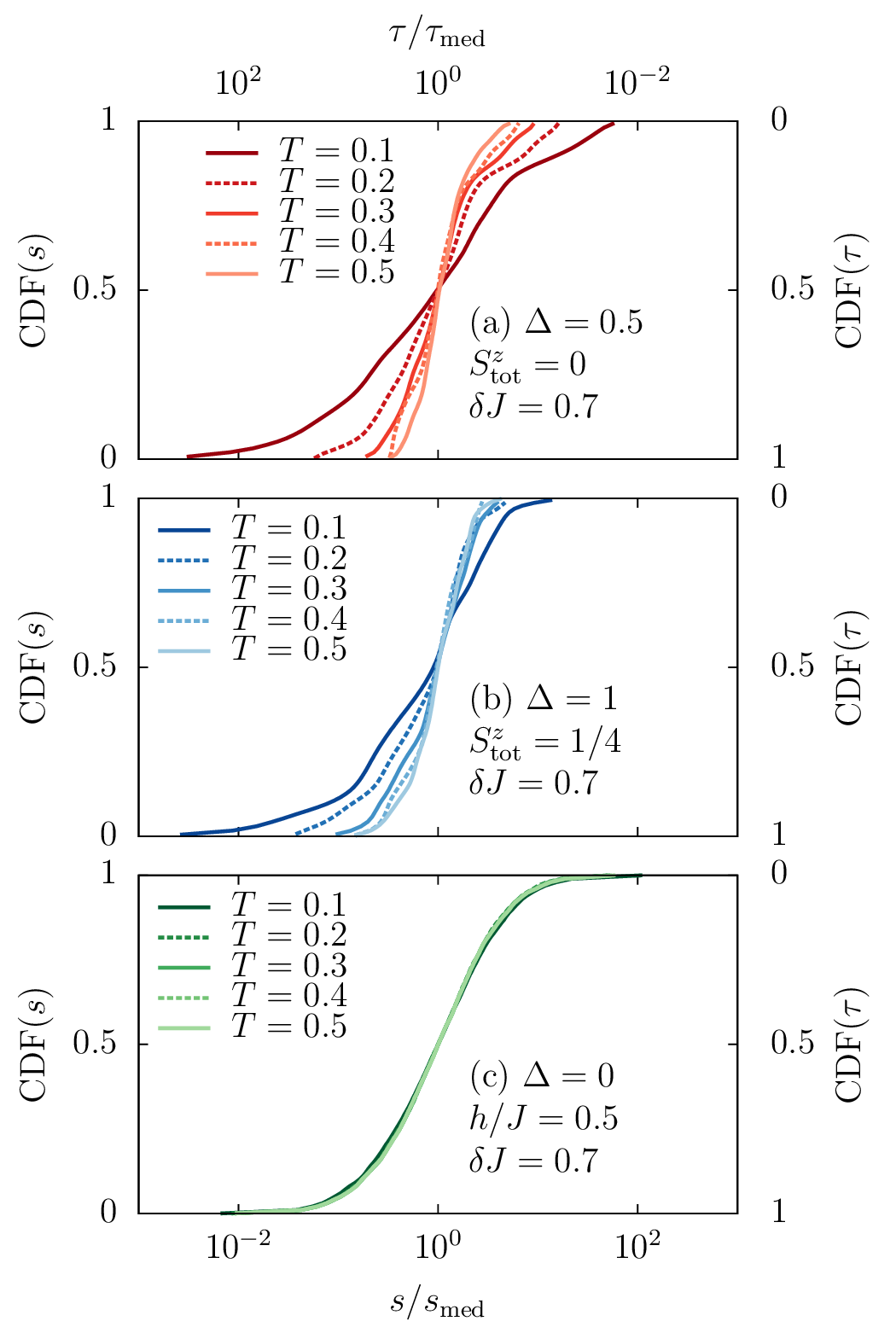}
\caption{(Color online) Cumulative distribution function of relaxation
rates $s$ and times $\tau=1/s$. Shown are FTD--DMRG results for
various $T \leq 0.5$ and: (a) for $\Delta=0.5$,
$S^z_{\mathrm{tot}}=0$ and $\delta J=0.7$, (b) for $\Delta=1$,
$S^z_{\mathrm{tot}}=1/4$ and $\delta J=0.7$, while (c) are
diagonalization results for $\Delta=0$, $\delta J=0.7$ and $h/J=0.5$.}
\label{cdf_supp}
\end{figure}

In Fig.~\ref{cdf_supp} we show that our main conclusions stay valid
also in a more general case, such as for $\Delta=0.5<1$ and for finite
magnetization ($S_{tot}^z=1/4$), where considerable
$T$ dependence of distribution with large spread is observed. In the
last panel of Fig.~\ref{cdf_supp} we show that the distribution for
noninteracting case (XX model) stays $T$ independent even in a finite
magnetic field $h$ or for finite magnetization.
 
\section{\rom{4.} Wavevector resolved spin structure factor}
Looking at the diverging uniform ($q=0$) susceptibility $\chi^0(T)$ as $T\to 0$
(Fig.~\ref{chi}) intuitively suggests large low--$q$ response and in
turn increasing contribution of $S(q\sim0,\omega)$ to the spin
relaxation rate $1/T_1$ as $T\to 0$. This is not what is observed,
since we see no increase of $1/\tau_0$ (Fig.~4b in the main text) as
$T\to 0$, but instead $1/\tau_0$ decreases with decreasing $T$, which
is in agreement also with experimental data
(Ref.~\cite{s_shiroka2011}, Fig.~3a). 

This dichotomy
can be partly understood by exploring the connection between static uniform
spin susceptibility $\chi^0(T)$ with the static spin structure factor 
(equal-time correlation)
$S(q)$, representing also the frequency integral of dynamical spin structure
factor $S(q)=\int _{-\infty}^{\infty}\mathrm{d}\omega\,S(q,\omega)$. The
connection 
$\chi^0(T)= S(q=0)/T$ together with the low--$T$ RG results (see Fig.~\ref{chi})
$\chi^0(T)=A/[T \ln^2(J_\textrm{max}/T)]$ leads to $S(q=0)=A/[
\ln^2(J_\textrm{max}/T)]$. This shows that $S(q=0)$ goes to 0 as $T\to 0$ and
is not diverging, rather its slow logarithmic approach to $0$ (in
contrast to linear in $T$ decrease for homogeneous system).
This is in agreement with results in Fig.~15 in Ref.~\cite{s_hoyos2007}, which
show that $S(q)$ at $T=0$ goes to 0 as $q\to 0$ and is only slightly
increased by randomness for $q\sim 0$.

\begin{figure}[!ht]
\includegraphics[width=1.0\columnwidth]{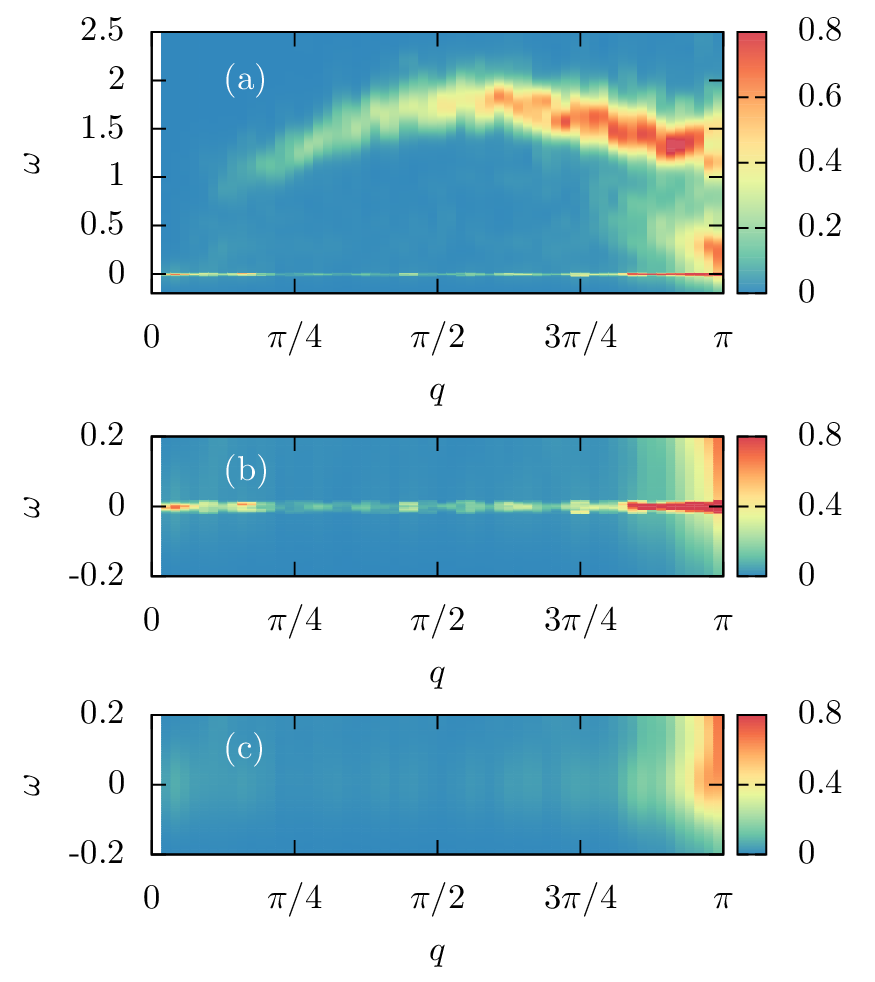}
\caption{(Color online) (a) Wavevector resolved $S(q,\omega)$ for $\delta J= 0.8$ and $T=0.1$ showing
much larger weight at low--$\omega$ at $q\sim \pi$. 
Panel (b) show $S(q,\omega)$ for low $\omega$ where the
non--dispersive $\omega=0$ delta peak is seen. The width
of the peak is due to thew broadening used in the
presentation. Panel (c) show low--$\omega$ 
$S(q,\omega)$ with removed $\omega=0$ delta peak.}
\label{s_fig_sqw}
\end{figure}

Another remarkable property of RHC can be seen in the
difference between dynamic ($\omega > 0$) properties, e.g. $S(q,\omega\to
0)$, and strictly $\omega=0$ contribution.
This is shown in the lower panel of Fig.~\ref{s_fig_sqw}, which
shows that $S(q,\omega)$ is singular at $\omega=0$ since it
consists besides the continuous background (regular part) 
also of a distinct delta peak at $\omega=0$ (see Fig.~\ref{s_fig_sqw}b). 
This peak is non--dispersive and is the signature of non--ergodicity in the 
random system (absence of diffusion) at least for low--$T$. Similar
peak is observed even in a random non--interacting
electron system at finite--$T$ (not presented). In our analysis of spin
relaxation for which $\omega \to 0$ is relevant, the $\omega=0$ peak
was excluded (see Fig.~\ref{s_fig_sqw}c and also Section \rom{1} above).
Distinction between strictly $\omega=0$ and $\omega>0$ properties 
can be traced back to the difference between diagonal elements, e.g. $\langle
(S^z_{tot})^2\rangle$ determining $\chi^0$ with only total spin $S>0$
(triplet) states contributing, and non--diagonal elements describing the
transitions between, e.g. singlet and triplet states, which are relevant
for dynamical ($\omega>0$) properties.

\subsection{\rom{4}a. Dominance of wavevectors $q\sim \pi$ in relaxation rate}
By approximating NMR relaxation rate $1/T_1$ with local
$S_\textrm{loc}(\omega\to 0)$, we neglected the effect of form factors
($A_\alpha^2(q)$, Eq.~2 in the main text),
which is a good approximation since in the regime of our calculations
the main contribution comes only from $q\sim \pi$. To show this, we
present $S(q,\omega)$ in Fig.~\ref{s_fig_sqw}, where it is evident
that the main contribution at $\omega\to0$ comes from $q\sim \pi$. This
stays valid even in the low--$T$ regime, where $\chi^0(T)$ is already
increased due to renormalization of $J$-s as observed in the RG flow. 
Randomness does in fact slightly reduces the contribution of $q\sim
\pi$ and slightly increases the contribution of $q\sim 0$, but the
transfer of weight is much too small to make $q\sim 0$ dominant. This
is in agreement with finding for $S(q)$ shown in Fig.~15 in
Ref.~\cite{s_hoyos2007}, where even for very large randomness and $T=0$
the main contribution stays at $q\sim \pi$ similarly to the homogeneous
system \cite{s_sandvik1995}. 
It also agrees with experimental observation and direct statement of the
authors \cite{s_shiroka2011}, that there in no indication of important
$q\sim 0$ contributions as, e.g., the d.c. field dependence of $1/T_1$, being
indication of the absence of (anomalous) low--$q$ (diffusion)
contribution. 

\subsection{\rom{4}b. Long--wavelength contributions}
Our analysis of local $s$ was based on assumption that there
is no singular contribution emerging from long--wavelength $q \to 0$
physics. Indeed, all our available data for $S^{zz}(q,\omega \to 0)$
for AFM RHC confirm that the dominant regime at low--$T$
is $q \sim \pi$. Still, $q \to 0$ regime needs further attention since it
can lead at $T>J$ to a divergent $S_{nn}(\omega \to 0) \propto
1/\omega^\alpha$ either from the propagation (prevented by randomness
in the RHC) in the homogeneous XX chain \cite{s_naef1999} (with $\alpha
\to 0$) or even more as the consequence of the spin diffusion 
\cite{s_sirker2009} ($\alpha =1/2$). The latter can be realized at $T>0$
but vanishes at $T \to 0$ within the RHC
\cite{s_theodorou1976,s_motrunich2001}. Our results so far indicate that
in spite of possible $T>0$ diffusion its contribution to
$S_{nn}(\omega \to 0)$ is unresolvable for reachable systems, as
follows also from NMR experiments \cite{s_shiroka2011} where it can be
directly tested via the magnetic field dependence of $T_1$.

\section{\rom{5.} Binary disorder distribution}
Concrete realization of a random system in Ref.~\cite{s_shiroka2011},
namely \ce{BaCu_{2}(Si_{0.5}Ge_{0.5})_{2}O_{7}}, has binary disorder
distribution with two exchange couplings,
$J_{\text{Si}}=280\,[K]$ and $J_{\text{Ge}}= 580\,[K]$, which is in
contrast with our continuous disorder distribution model, motivated by a
usual theoretical reference. Therefore the question arises, how
different are the results for the binary distribution from our results. Here we argue
that the results are qualitatively and even
quantitatively very similar for both distributions. This was realized
already by J.~E.~Hirsch \cite{s_hirsch1980}, who showed that arbitrary
disorder distribution lead to the similar low--$T$ behaviour. 

\begin{figure}[!ht]
\includegraphics[width=1.0\columnwidth]{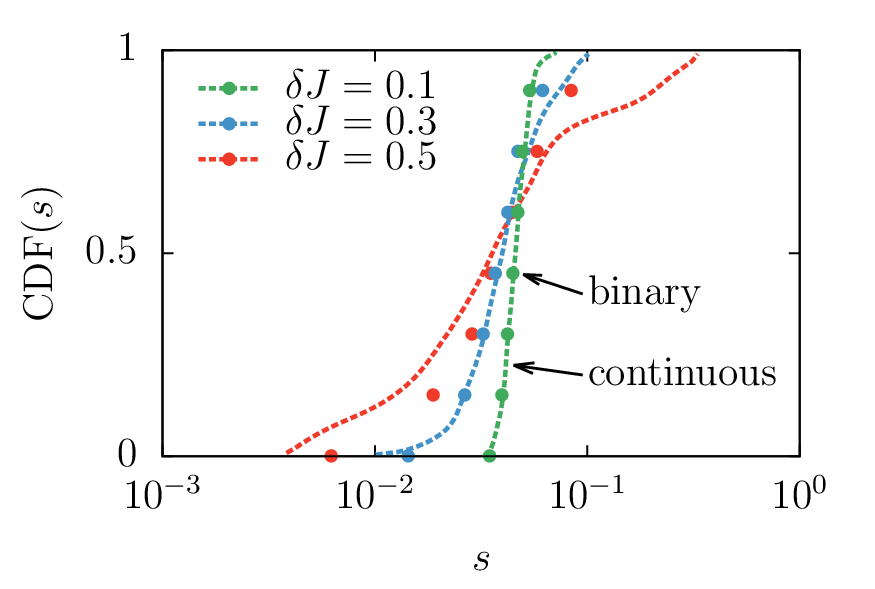}
\caption{(Color online) Comparison of CDF-s for continuous and binary
disorder distribution. Both distributions give almost the same
CDF for small disorder, while for larger disorder they show only
small quantitative difference. Note that the $x$--axis is not rescaled and
both distributions would therefore give almost the same $1/\tau_0$.
Plots are for $L=80$, $T=0.3$ and obtained with $N_r=10^3$ realizations.}
\label{s_fig_bin}
\end{figure}

To demonstrate the effect of binary distribution we show in
Fig.~\ref{s_fig_bin} the comparison of relaxation rate CDF-s for continuous
disorder distribution with the ones for binary disorder
distribution with the same effective width. It is seen that the 
difference is small and largest for 
strongest disorder, where it still remains only quantitative, while for low
disorder CDF-s are essentially the same for both disorder
distributions. Therefore our results obtained with continuous distribution
can easily be compared with measurements and they indeed agree
qualitatively and to some extend even quantitatively with them (see
main text).

\section{\rom{6.} Comments on stretched exponential}
Using phenomenological stretched exponential form to fit experimental
data on magnetization relaxation seems to be a common practice, which
can be attributed to the fact that stretched exponential form can capture
anomalously long tails in the distribution of the relaxation rates
(normal distribution can not) and is at the same time very convenient
for the fitting procedure. This immediately raises the question, how
good this form really is for the description of experimental data and
can it be motivated by some microscopic picture, e.g. model
Hamiltonian. 

\begin{figure}[!ht]
\includegraphics[width=1.0\columnwidth]{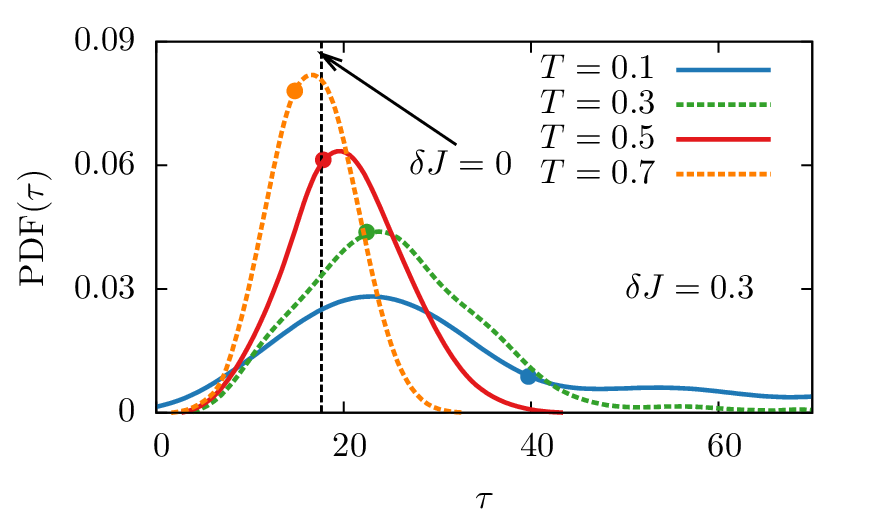}
\caption{(Color online) Probability distribution function PDF($\tau$)
for several temperatures and one randomness $\delta J=0.3$. These results should be
compared with the shape of experimental ones in Fig.~4 in
Ref.~\cite{s_shiroka2011}. Temperature evolution of the distribution
agrees with the experiment, as well as appearance of anomalously
long tails of the distribution at low--$T$. Points correspond to
fitted $\tau_0$ which shows similar $T$--dependence as experiment,
while dashed vertical line corresponds to homogeneous ($\delta J=0$)
model.}
\label{s_fig_pdf}
\end{figure}

In Fig.~\ref{s_fig_pdf} we show our RHC model results for relaxation
time distributions (PDF-s), which shows several important features when
compared to the experimentally suggested stretched exponential forms
(see Fig.~4 in Ref.~\cite{s_shiroka2011}). First one can see that the
$T$ evolution is similar to the experimental one and
more importantly at low--$T$ anomalously long tails (or large spread) in PDF appear,
which can be captured with stretched exponential form and not with,
e.g., normal (Gaussian) distribution. This could be the reason for the
success of the stretched exponential form in the fitting procedures
and its phenomenological description of
experimental data.

\begin{figure}[!ht]
\includegraphics[width=1.0\columnwidth]{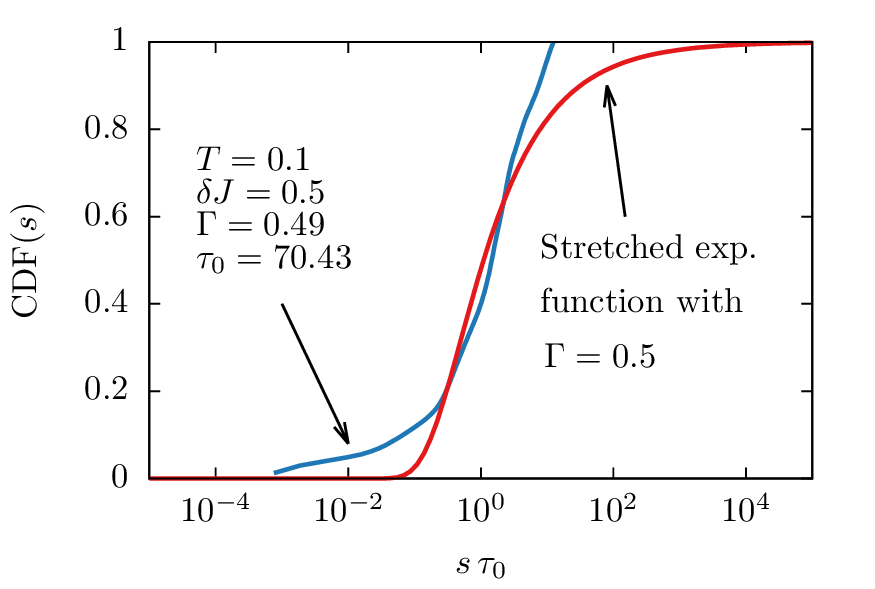}
\caption{(Color online)
One of more critical comparisons of our calculated CDF (blue line) to approximated
stretched exponential form (red line) for $\Gamma \simeq 0.5$, for
which stretched exponential distribution has particularly simple
analytical form \cite{s_lindsey1980,s_johnston2006}.}
\label{s_fig_sea}
\end{figure}

However, the description of our RHC model results with stretched
exponential is not perfect as can be expected, and the most obvious
deviations can be found in the long tails. E.g., for few specific value of $\Gamma$
analytical form of PDF-s is know \cite{s_lindsey1980,s_johnston2006}
and for $\Gamma=0.5$ has a form 
\begin{equation}
\mathrm{PDF}_{\Gamma=0.5}(s)=\frac{\exp\left[-1/\left(4\tau_0s\right)\right]}{\sqrt{4\pi\tau_0s^3}}\,.\label{pdfan}
\end{equation}
On Fig.~\ref{s_fig_sea} we compare our numerical result with
$\Gamma=0.49$ (for $T=0.1$ and $\delta J=0.5$) with stretched
exponential CDF obtained from Eq.~\eqref{pdfan},
$\mathrm{CDF}_{\Gamma=0.5}(s)=\mathrm{Erf}[1/(2\sqrt{s\,\tau_0})]$.
We observe that the RHC model predicts longer (shorter) tails in the PDF for smaller
(larger) $s$ than stretched exponential form. This is in turn 
reflected in the corresponding time dependent magnetization relaxation
function (being Laplace transform of PDF) directly probed by experiment. One
could, for example, from 
our PDF-s propose a new form (instead of stretched exponential) by
approximating PDF-s with some function and performing its Laplace transform. This
is however not trivial and we leave it as a motivation for future
work. In this way obtained form is expected to describe experimental
data better than stretched exponential, although differences might be
small and experimental data with higher resolution might be
needed. 

\end{document}